\documentclass[aps,prl,twocolumn,superscriptaddress,english]{revtex4}
\usepackage{amssymb}
\usepackage{graphicx}
\usepackage{amsmath}
\usepackage{soul}
\usepackage{amsthm}
\usepackage{amsfonts}
\usepackage[T1]{fontenc}
\usepackage[latin9]{inputenc}
\usepackage{array}
\usepackage{multirow}
\usepackage{color}
\usepackage{esint}
\usepackage{bm}
\usepackage{color}
\usepackage{bbm}
\usepackage{hyperref}
\usepackage{babel}
\usepackage{titlesec}

\begin{document}

\global\long\def\id{\mathbbm{1}}
\global\long\def\ui{\mathbbm{i}}
\global\long\def\ud{\mathrm{d}}

\title{Quantum phase with coexisting localized, extended, and critical zones}
\author{Yucheng Wang}
\affiliation{Shenzhen Institute for Quantum Science and Engineering,
Southern University of Science and Technology, Shenzhen 518055, China}
\affiliation{International Quantum Academy, Shenzhen 518048, China}
\affiliation{Guangdong Provincial Key Laboratory of Quantum Science and Engineering, Southern University of Science and Technology, Shenzhen 518055, China}
\author{Long Zhang}
\affiliation{School of Physics, Huazhong University of Science and Technology, Wuhan 430074, China}
\author{Wei Sun}
\affiliation{Department of Physics, Southern University of Science and Technology, Shenzhen 518055, China}
\author{Ting-Fung Jeffrey Poon}
\affiliation{International Center for Quantum Materials, School of Physics, Peking University, Beijing 100871, China}
\affiliation{Collaborative Innovation Center of Quantum Matter, Beijing 100871, China}
\author{Xiong-Jun Liu}
\thanks{Corresponding author: xiongjunliu@pku.edu.cn}
\affiliation{International Center for Quantum Materials, School of Physics, Peking University, Beijing 100871, China}
\affiliation{Collaborative Innovation Center of Quantum Matter, Beijing 100871, China}
\affiliation{Shenzhen Institute for Quantum Science and Engineering,
Southern University of Science and Technology, Shenzhen 518055, China}
\begin{abstract}
Conventionally a mobility edge (ME) marks a critical energy that separates two different transport zones where all states are extended and localized, respectively. Here we propose a novel quasiperiodic spin-orbit coupled lattice model with experimental feasibility to realize a fundamentally new quantum phase with three coexisting energy-dependent zones, i.e., the extended, critical and localized zones, and uncover the underlying generic mechanism for the occurrence of the new quantum phase. Accordingly, this phase exhibits new types of MEs which separate the extended states from critical ones, and the localized states from critical ones, respectively. We introduce the diagnostic quantities to characterize and distinguish the different zones, and show that the predicted phase can be detected by measuring the fractal dimension or conductivities. The experimental realization is also proposed and studied. This work extends the concept of ME and enriches the quantum phases in disordered systems, 
which sheds light on searching for new localization and critical phenomena with novel transport and thermoelectric effects. 
\end{abstract}
\maketitle

\begin{figure}
\hspace*{0.2cm}
\centering
\includegraphics[width=0.47\textwidth]{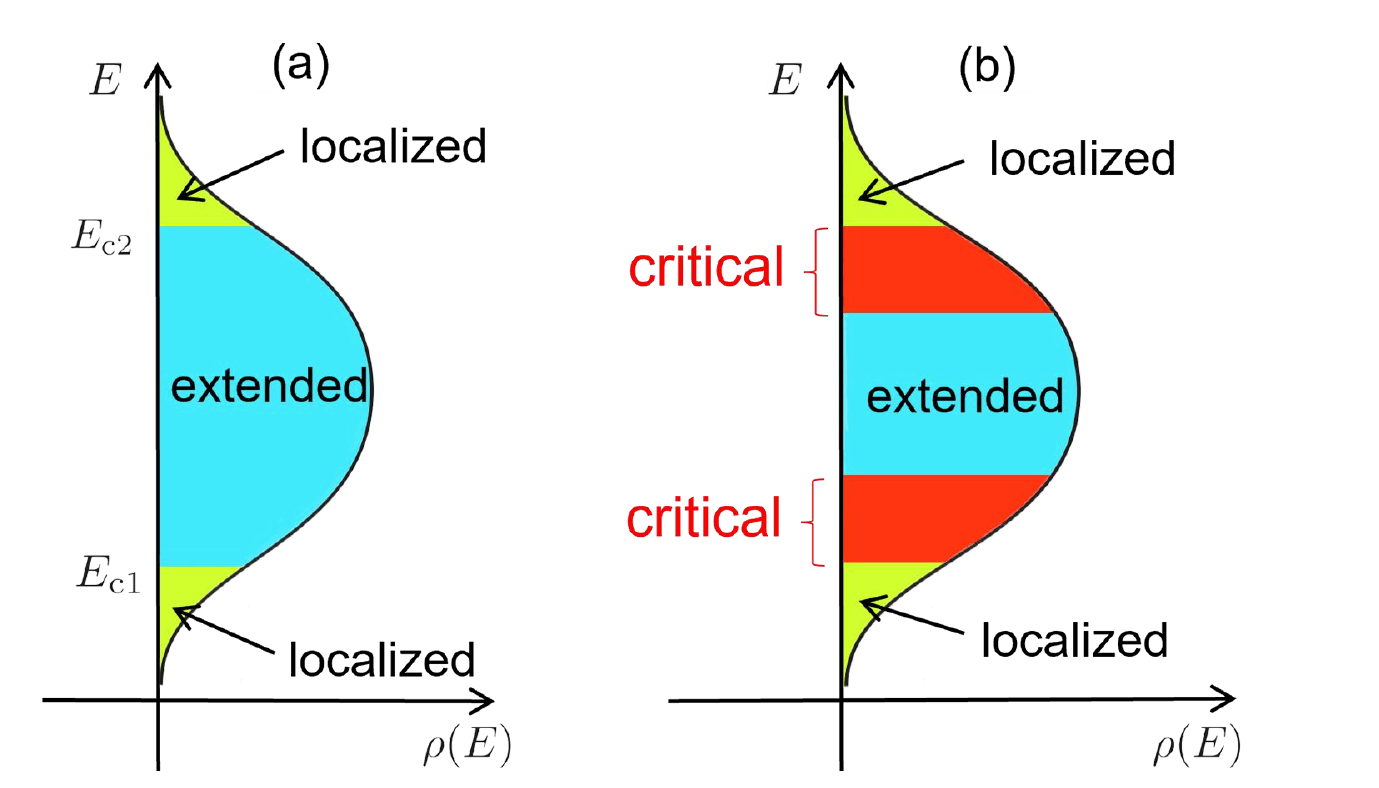}
\caption{\label{01}
Schematic figures of the density of states $\rho(E)$ versus the energy $E$ for a system with (a) conventional MEs separating the extended and localized states, and (b) three different energy-dependent transport zones (extended, critical and localized) separated by the new type of MEs.}
\end{figure}

{\em Introduction.---}Disordered potential is ubiquitous in quantum materials and, as Anderson envisioned, can induce exponentially localized electronic wave-functions when the randomness is sufficiently strong~\cite{Anderson1958}. This implies the existence of disorder-driven Anderson transition (AT) between extended and localized phases~\cite{RMP1,RMP2,Kramer}. The AT may also appear versus eigen-energies in a single phase in the moderate disorder strength regime, namely, a part of the states are localized and the rest are extended, with the localized and extended zones being separated by critical energies called mobility edges (MEs)~\cite{RMP1}. A typical situation is illustrated in Fig.~\ref{01}(a), with $E_{c1}$ and $E_{c2}$ denoting the MEs, where the states in band tails are localized, while those in the center remain extended. 
The AT can also be obtained in quasiperiodic lattices
~\cite{AA,Xie1988,Biddle2009,Biddle,Ganeshan2015,XDeng,Saha2019,XLi,HYao,Wang1,XPLi,TongLiu,Wang2}, in which the quasiperiodic potential has important difference from the random disorders. In particular, the AT and ME can exist in one-dimensional (1D) quasiperiodic systems~\cite{AA,Xie1988,Biddle2009,Biddle,Ganeshan2015,XDeng,Saha2019,XLi,HYao,Wang1,XPLi,TongLiu,Wang2}, but in disordered system they only exist in dimension higher than two according to the scaling theory~\cite{Thouless,Anderson1979}. 
Quasiperiodic systems can be realized in ultracold gases by using two optical lattices with incommensurate wavelengths~\cite{Roati}, with which the AT and ME have been observed recently in experiments~\cite{Roati,Bloch1,Gadway1,Gadway2}.

The quasiperiodic systems can also host a third type of phases called critical phases~\cite{Kohmoto,Takada,Fliu,JWang}, 
which are extended but non-ergodic~\cite{XDeng,XPLi,Kravtsov2015,Kravtsov2018,Whitlock2019,Das2022}, and are fundamentally different from the extended and localized phases in the spectral statistics~\cite{Geisel,Jitomirskaya,explain0}, multifractal properties of wave functions~\cite{Halsey,Mirlin}, and dynamical evolutions~\cite{Hiramoto,Ketzmerick}.
In the presence of interactions, the single-particle critical phase can become the many-body critical phase~\cite{Wang2,Wang3,BoYan}. The critical phase is a fertile ground to explore various important physics such as the non-ergodic physics, critical transport behavior, AT, and thermal-nonthermal transitions~\cite{Huse1,Huse2,Kulkarni2018,Serbyn2019}. An important open question is whether a finite critical zone exists between the extended and localized zones in a single phase?
Namely, is there a quantum phase with coexisting three fundamentally different zones with all eigenstates in one zone having the same properties, as illustrated in Fig.~\ref{01}(b)?
Presumably such a novel phase, if existing, shall host novel new quantum physics including new type of energy-dependent transport features (ballistic, normal diffusive, and localized), new quantum dynamics, and new thermoelectric response~\cite{thermal1,thermal2,thermal3,Whitney,Hatano,Saha1907,Goold}, which could be observed in experiment by tuning Fermi energy.
Therefore, to uncover a system with three different coexisting energy-dependent transport zones is of both fundamental importance and potential applications, but was not considered before. 

In this Letter, we predict a novel quantum phase with coexisting localized, extended, and critical zones in a 1D quasiperiodic model, which can be realized in current experiments, and uncover the underlying generic mechanism for the coexisting phase. This new phase exhibits two types of MEs separating the extended and localized zones from critical zones, respectively, 
with new diagnostic quantities being introduced to characterize this phase.

{\em Model and phase diagram.---}We propose the quasiperiodic optical Raman lattice model described by
\begin{equation}\label{Hsum}
H=H_t+H_{\rm SOC}+H_{Z},
\end{equation}
with
\begin{subequations}
\begin{eqnarray}
H_t=-t_0\sum_{\langle i,j\rangle}(c^{\dagger}_{i,\uparrow}c_{j,\uparrow}-c^{\dagger}_{i,\downarrow}c_{j,\downarrow}), \qquad \qquad \qquad\\
H_{\rm SOC}=t_{so} \sum_i(c^{\dagger}_{i,\uparrow}c_{i+1,\downarrow}-c^{\dagger}_{i,\uparrow}c_{i-1,\downarrow})+H.c., \\
H_{Z}=\sum_{i}\delta_i(n_{i,\uparrow}-n_{i,\downarrow})+ \lambda\sum_{i}\delta_i(n_{i,\uparrow}+n_{i,\downarrow}),
\end{eqnarray}
\end{subequations}
where $c_{j,\sigma}$($c^{\dagger}_{j,\sigma}$) is the annihilation (creation) operator for spin $\sigma=\uparrow,\downarrow$ at lattice site $j$ and $n_{j,\sigma}=c^{\dagger}_{j,\sigma}c_{j,\sigma}$ is the particle number operator, the term $H_t$ ($H_{\rm SOC}$) denotes the spin-conserved (spin-flip) hopping coupling between neighboring sites with strength $t_0$ ($t_{so}$), and the last term $H_Z$ includes spin-dependent and spin-independent quasiperiodic potentials of strengths $M_z$ and $\lambda M_z$, respectively, with $\delta_j=M_z\cos(2\pi\alpha j+\phi)$ and $\alpha$ ($\phi$) being an irrational number (a phase shift). This system with uniform Zeeman energy, i.e., $\delta_i=m_z$, and 
$\lambda=0$ gives the 1D AIII class topological insulator and is experimentally realized~\cite{XJLiu,BSong,BZWang,LZhang}.
For convenience, we set $t_0=1$ as the unit energy, $\phi=0$, and $\alpha=\lim_{m\rightarrow\infty}\frac{F_{m-1}}{F_m}=(\sqrt{5}-1)/2$, with $F_m$ being the Fibonacci numbers~\cite{Kohmoto1983,Fibonacci}. For finite system with size $L=F_m$ one takes $\alpha=\frac{F_{m-1}}{F_m}$ to introduce the periodic boundary condition.

\begin{figure}
\hspace*{-0.3cm}
\centering
\includegraphics[width=0.5\textwidth]{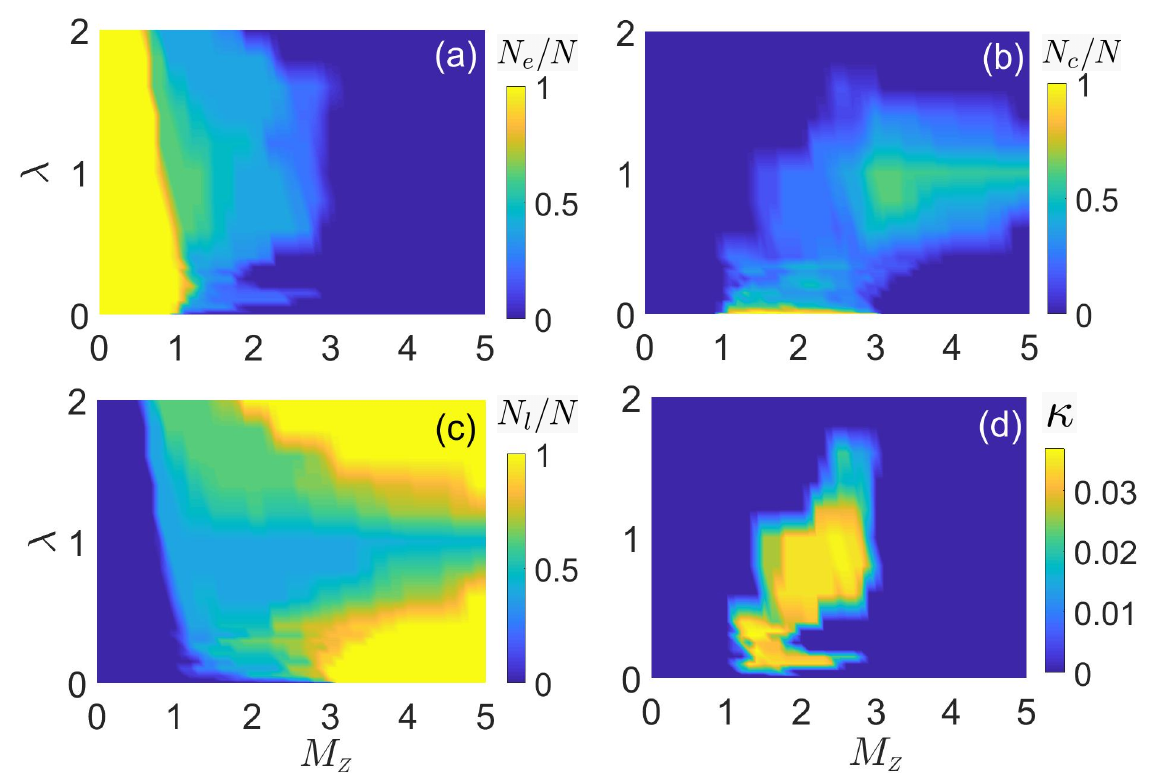}
\caption{\label{02}
(a) The extended fraction $N_e/N$, (b) The critical fraction $N_c/N$, (c) The localized fraction $N_l/N$, and (d) $\kappa$ as the function of $\lambda$ and $M_z$ with $t_{so}=0.5$.}
\end{figure}

To characterize the key physics of the above model, we define a characteristic quantity by
\begin{equation}\label{Gamma}
\kappa=\frac{N_e}{N}\times\frac{N_c}{N}\times\frac{N_l}{N},
\end{equation}
where $N_e$, $N_c$, $N_l$ and $N=2L$ are the numbers of the extended, critical, localized and total eigenstates, respectively, providing the diagnostic quantities of all different phases. 
The conventional ME separating localized and extended states corresponds to $N_eN_l/N^2 >0$ with $\kappa=0$, while $N_eN_c/N^2>0$ (or $N_cN_l/N^2>0$) characterizes a new type of ME separating critical states from extended (or localized) ones. The most nontrivial phase corresponds to $\kappa>0$ which characterizes the coexisting localized, extended, and critical zones. Fig.~\ref{02} show the $N_{e,c,l}/N$ and $\kappa$, which are calculated by comparing the fractal dimension (FD) with different sizes (see below).
When $\lambda=0$, this system hosts three distinct phases with solely extended ($N_e/N=1$), critical ($N_c/N=1$), or localized ($N_l/N=1$) eigenstates~\cite{Wang2}. However, with increasing $\lambda$ to be within a proper range, an unprecedented new phase emerges with three coexisting zones and $\kappa>0$ [Fig.~\ref{02}(d)]. 
This new phase occurs in a wide range of parameter about $0.02<\lambda<1.6$ for $t_{\rm so}=0.5$ [see Supplementary Material (SM)~\cite{SM}].
Thus the presence of both spin-independent and spin-dependent quasiperiodic potentials, with strength $\lambda$ being freely adjustable in experiment (see below), is terribly important for the coexisting phase, whose occurrence deeply stems
from breaking chiral symmetry of this incommensurate AIII class system~\cite{SM}. 
This key ingredient marks the 
fundamental difference in comparison with the previous study~\cite{Wang2}.

\begin{figure}
\hspace*{-0.1cm}
\centering
\includegraphics[width=0.5\textwidth]{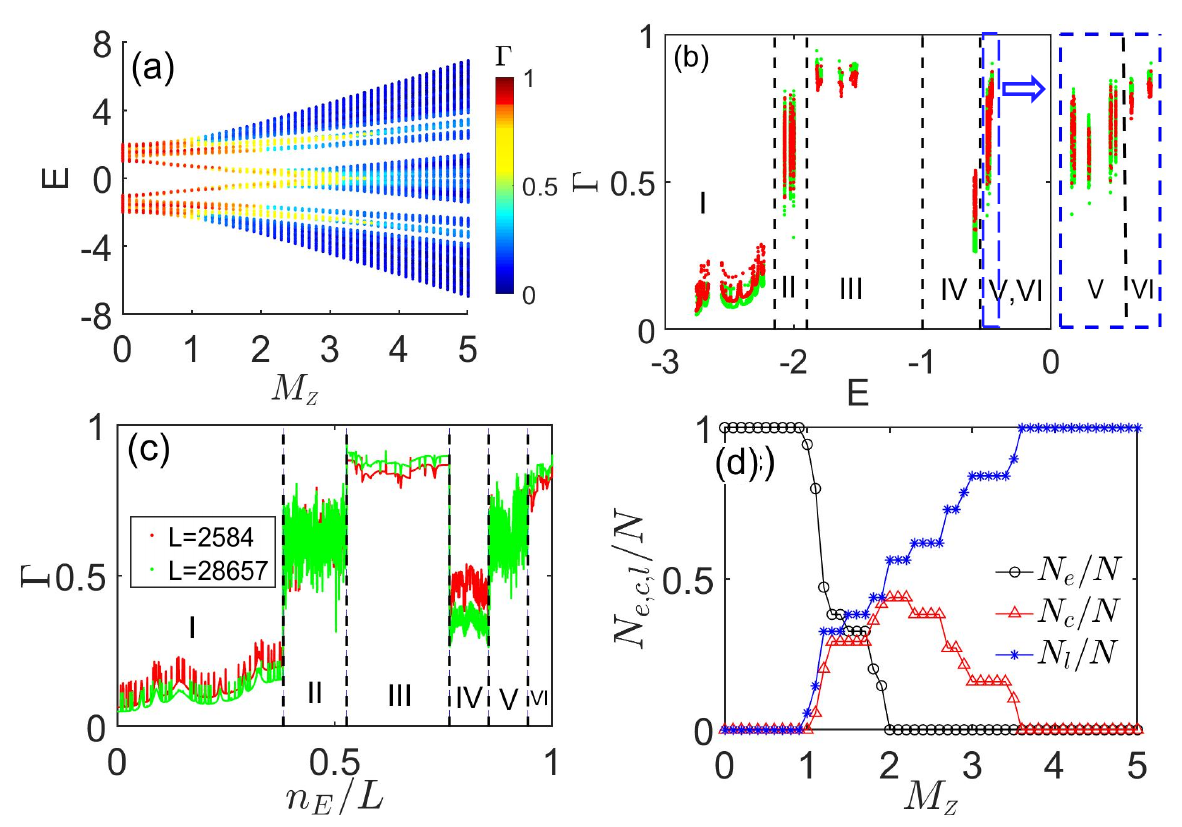}
\caption{\label{03}
(a) Fractal dimension $\Gamma$ of different eigenstates as the function of the corresponding eigenvalues $E$ and $M_z$ with size $L=F_{14}=610$. (b) $\Gamma$ of different eigenstates versus the corresponding eigenvalues $E$ for different sizes $L=F_{17}=2584$ (red) and $L=F_{22}=28657$ (green) with $M_z=1.5$. The right blue dashed frame is the enlarged part of the left blue dashed frame. (c) Compared to (b), the horizontal axis becomes $n_E/L$ with $n_E$ being the index of energy mode. (d) $N_e/N, N_c/N$ and $N_l/N$ as a function of $M_Z$. Other parameters are $t_{\rm so}=0.5$, $\lambda=1/3$.}
\end{figure}

To further confirm the existence of the new phase, we present a quantitative study with $\lambda=1/3$. The different types of states can be identified by the FD $\Gamma$, and for an arbitrary state $|\psi\rangle=\sum_{j,\sigma}^L u_{j,\sigma}c^{\dagger}_{j,\sigma}|0\rangle$, it is defined as
\begin{equation}\label{Gamma}
\Gamma=-\lim_{L\rightarrow\infty}\ln(IPR)/\ln L,
\end{equation}
where $IPR=\sum_{j,\sigma}u^4_{j,\sigma}$ is the inverse participation ratio~\cite{RMP1}. The FD tends to $1$ and $0$ for the extended and localized states, respectively, while $0<\Gamma<1$ for the critical states.
Fig.~\ref{03}(a) shows $\Gamma$ of different eigenstates as a function of $M_z$ and the corresponding eigenvalues $E$, from which one identifies the phase regions with coexisting three or two different types of eigenstates separated by MEs. Particularly, we fix $M_z=1.5$, which belongs to the coexisting region of extended, critical and localized zones, and show $\Gamma$ of different eigenstates in Fig.~\ref{03}(b) as a function of the corresponding eigenvalues for different sizes. For convenience, we present the region with $E<0$, which is symmetric to that with $E>0$. One can observe that $\Gamma$ tends to $0$ for all states in zones I and IV 
with increasing the system size, implying that they are localized, while $\Gamma$ tends to $1$
for all states in zones III and VI 
when increasing the size, showing that all states are extended. In contrast, in zones II and V, the FD $\Gamma$ is clearly different from $0$ and $1$, and peculiarly, is almost independent of the system size, meaning that all states are critical. To distinguish different zones more easily, one can use the ratio of the index of energy mode $n_E$ to $L$ as the the horizontal axis, instead of the eigenvalue $E$, as shown in Fig.~\ref{03}(c). We see the distinguishable zones and the sudden change between different zones. More rigorously, in SM~\cite{SM} and the following discussion about the momentum-space distributions, we study the scaling behavior of different zones and further confirm the coexisting phase. After determining the energy windows of extended, critical and localized zones, one can easily compute numerically 
$N_e/N, N_c/N, N_l/N$, which are size-independent~\cite{SM} and can also be directly obtained from Fig.~\ref{03}(c)~\cite{explain}. Fig.~\ref{03}(d) displays the three quantities, for $1<M_z<2$, which are all larger than $0$, signifying the phase with three coexisting zones.

Before proceeding we point out the generic underlying mechanism of the coexisting phase, with which we show that this phase can exist in broad range of quasiperiodic systems. As detailed in SM~\cite{SM}, the current model with nonzero $\lambda$ but zero spin-orbit coupling $t_{\rm so}=0$ has ME separating extended and localized states. On the other hand, the model with $\lambda=0$ hosts critical phase~\cite{Wang2}. The appropriate combination of the way to induce mobility edge and the way to induce critical phase give rise to the coexisting phase. There are various schemes to obtain mobility edges, such as, the quasiperiodic systems with the next-nearest-neighbor~\cite{Biddle2009}, exponential~\cite{Biddle} or power-law hopping~\cite{XDeng} term, or multiple potentials~\cite{XLi,HYao}. Combining either of them with the way to induce critical phase can yield the coexisting phase. As we confirm in the SM, instead of considering spin-independent quasiperiodic potential (i.e. $\lambda=0$), when a next-nearest-neighbor hopping term is added, or if the long-range hopping is present, the coexisting phase can also be readily obtained. These results show that this coexisting phase can be naturally realized in various quasiperiodic models in real condensed matter and cold atom systems.

\begin{figure}
\hspace*{-0.1cm}
\centering
\includegraphics[width=0.5\textwidth]{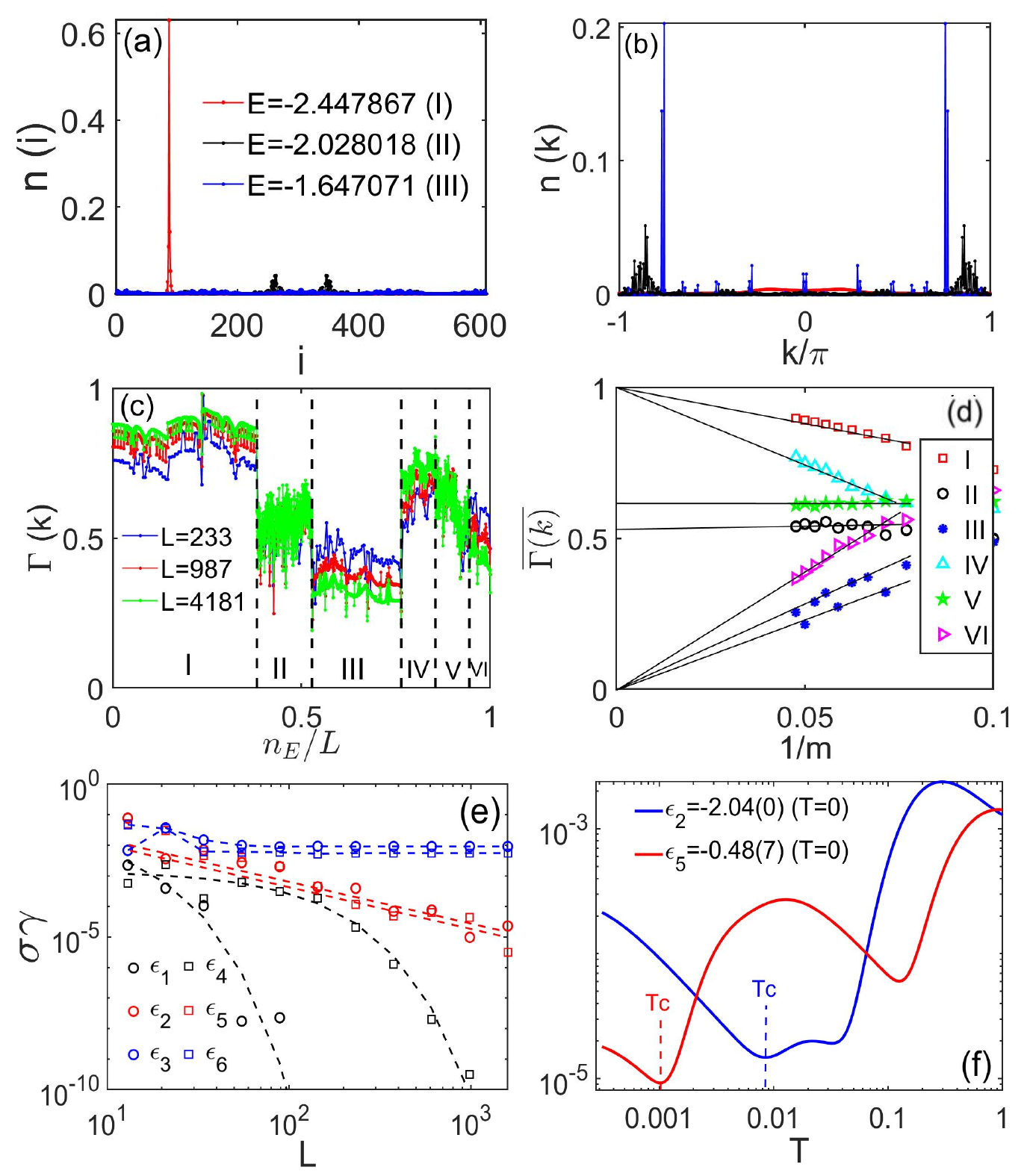}
\caption{\label{04}
The typical distribution of the eigenstates in zone I (localized zone), II (critical zone), and III (extended zone) in (a) real space and (b) momentum space. (c) fractal dimension $\Gamma$ in momentum space for each eigenstate at three different sizes.
(d)$\overline{\Gamma(k)}$ as a function of $1/m$ for different zones, where $m$ are the Fibonacci indices.  (e) $\sigma\gamma$ versus $L$ at zero temperature. The Fermi surfaces $\epsilon_{1-6}$ in zones I-VI are chosen at $-2.67(4), -2.07(8), -1.79(8), -0.57(7), -0.48(8)$ and $-0.45(7)$, respectively. (f) $\sigma\gamma$ versus temperature with $\alpha=55/89$. Fermi surfaces in zones II and V are chosen at $\epsilon_2=-2.04(0)$ and $\epsilon_5=-0.48(7)$, respectively. Other parameters are $t_{so}=0.5$, $\lambda=1/3$ and $M_Z=1.5$.}
\end{figure}

{\em Detecting three different zones.---}The detection of the three different characteristic zones can be achieved by contrasting the distribution of the wave-functions in real and momentum spaces. The momentum distribution is defined as $n(k)=n_{\uparrow}(k)+n_{\downarrow}(k)$ for an eigenstate $|\psi\rangle$, where $n_{\sigma}(k)=\langle\psi|c^{\dagger}_{\sigma, k}c_{\sigma, k}|\psi\rangle$ and $c_{\sigma,k}=(1/\sqrt{L})\sum_{l}e^{ikl}c_{\sigma,l}.$
The distributions of localized (extended) states are localized (extended) in real space but extended (localized) in momentum space [see the red (blue) lines in Fig.~\ref{04}(a) and (b)]. 
However, for the critical states the distributions are delocalized and non-ergodic in both real and momentum spaces [see the black lines in Fig.~\ref{04}(a) and (b)]. These qualitatively distinct features characterize and can be applied to detect different zones.

The momentum distribution of a quantum gas can be readily measured with time of flight imaging.
Analogue to the FD in real space, we introduce the IPR in momentum space $IPR(k)=\sum_{k_m,\sigma}n_{k_m,\sigma}^2$ with $k_m=\pi\frac{m}{L}$ ($m=-L,-L+1,\dots,L-1$) and define the FD $\Gamma(k)=-\lim_{L\rightarrow\infty}\ln(IPR(k))/\ln L$.
But when varying the system size, the magnitudes and number of the eigen-energies may change accordingly. 
To avoid the potential challenge in contrasting $\Gamma(k)$ for a fixed state at different sizes, it is convenient to observe the average value over the eigenstates in a single zone, 
and accordingly we define the zone FD
\begin{equation}\label{aeta}
\overline{\Gamma(k)}=\frac{1}{N_s}\sum_{same\: zone}\Gamma(k),
\end{equation}
where $N_s$ is the number of states of the zone, i.e., $\overline{\Gamma(k)}$ is obtained by calculating the average $\Gamma(k)$ of all states in the same zone. We note that since all eigenstates in the same zone have the same properties, the mean FD in an arbitrary small sub-zone of a zone can also be similarly defined and will show the same scaling behavior with the zone~\cite{explain2}. Fig.~\ref{04} (c) and (d) respectively show the $\Gamma(k)$ and $\overline{\Gamma(k)}$, we see that they extrapolates to $0$ in zones III and VI, to $1$ in zones I and IV, while to the values far from $0$ and $1$ in zones II and V, which confirm that the corresponding states in these zones are extended, localized and critical in momentum space, respectively.


Ultracold atom experiments consisting of a channel connected two reservoirs have been used to investigate and detect interesting transport behaviors~\cite{Chien2015,Brantut2012,Brantut2013,Krinner2013,Krinner2015}, which can qualitatively distinguish the three different zones.
We consider the conductivity $\sigma$ based on the Kubo formula in the dissipationless limit $\gamma\rightarrow 0$~\cite{Mahan,Mitscherling2020},
\begin{equation}
\sigma = \frac{1}{2\gamma h}\times \int dk \sum_b \left(-\frac{df_{FD}(E_b(k) - \mu)}{dE_b(k)}\right) (\frac{dE_b(k)}{dk})^2
\end{equation}
where $f_{FD}$ is the Fermi-Dirac distribution, $E_b$ is the energy of the $b$th band, $\mu$ is the chemical potential and $\gamma$ is the scattering rate~\cite{gamma} assumed to be momentum and band independent.
Fig.~\ref{04} (e) shows the conductivity in the zero temperature. We see that for Fermi surface in the extended zones, the conductivity is independent of system size ($\sigma\sim L^0$), while in the localized zones, conductivity decreases exponentially with the system size ($\sigma\sim e^{-L}$). The conductivity in the critical zones decays in a power-law fashion ($\sigma\sim L^{-p}$, where $p$ is a parameter). The different scaling behaviors of the conductance through varying the Fermi energy clearly show the new types of MEs.
With increasing the temperature, the transport behaviors of the critical zones II and V are showed in Fig.~\ref{04} (f). The transport behavior is more sensitive to temperatures when the distance between the Fermi surface (labelled as $\epsilon_F$) and the minimum value of the above neighbor extended zone (labelled as $E^{e}_{min}$) is smaller. There is a temperature $T_c$ ($T_c\propto (E^{e}_{min}-\epsilon_F)/k_b$ with $k_b$ being the Boltzmann's constant) corresponding to that the transport effect of the particles that enter the extended zones becomes non-ignorable and thus, the conductivity rapidly increases. $T_c$ for zone V is smaller, indicating that both its energy range and the gap between zone V and VI are narrower. To conclude, this new phase shows the abundant and new transport features.

{\em Proposal of experimental realization.---}Here we propose to realize the Hamiltonian~\eqref{Hsum} in cold atoms by employing the optical Raman lattice scheme.
We choose two hyperfine states $|F,m_F\rangle$ and $|F',m'_F\rangle$ to construct the spin-1/2 system, apply a standard spin-independent lattice potential to generate spin-conserved hopping term ($H_t$), and a Raman coupling potential to induce spin-flip hopping term ($H_{\rm SOC}$)~\cite{Wang2,XJLiu,BSong,BZWang,LZhang}. The incommensurate potential term ($H_Z$) can be generated from scalar and vector potentials induced by two counter-propagating lights with proper polarizations. 
The realized Hamiltonian reads~\cite{SM}
\begin{equation}\label{Htotal}
H=\left[\frac{k_z^2}{2m}+V_1(z)+\lambda V_2\right]\otimes\id+{\cal M}(z)\sigma_x+\left[V_2(z)+\frac{\delta}{2}\right]\sigma_z,
\end{equation}
where $V_1(z)=V_p\cos^2(k_1z)$ is the primary lattice, $V_2(z)=\frac{V_s}{2}\cos(2{k}_2z+\phi)$ denotes the secondary spin-dependent weak lattice, giving an incommensurate Zeeman potential with the irrational number $\alpha=k_2/k_1$, ${\cal M}(z)=M_0\cos(k_1z)$ is the Raman coupling potential, and $\delta$ is the two-photon detuning. In the SM~\cite{SM}, we show that this continuous Hamiltonian (\ref{Htotal}) indeed leads to the model Hamiltonian described by Eq.~(\ref{Hsum}).

Here we mainly discuss the key parameter $\lambda$. For chosen magnetic quantum numbers $m_F$, $m'_F$, $\lambda$ has a minimal value $\lambda_{\rm min}$ when the polarization is mutually perpendicular~\cite{SM}.
In contrast, if the two lights have the same and linear polarization, incommensurate lattice becomes fully spin-independent, corresponding to that $\lambda\to\infty$.
Hence, $\lambda$ can vary within the range $\lambda_{\rm min}\leq \lambda<\infty$ by tuning the light polarization.
We consider the mutually perpendicular case, i.e. $\lambda=\lambda_{\rm min}$ and obtain~\cite{SM}
\begin{align}
\lambda_{\rm min}=\frac{g_F m_F+g_{F'}m_F'}{g_F m_F-g_{F'}m_F'},
\end{align}
where $g_{F}$ ($g_{F'}$) is the Land\'e factor for the $F$ ($F'$) hyperfine states.
In particular, when $g_{F}=-g_{F'}$, the result reduces to $\lambda_{\rm min}=(m_F-m_{F'})/(m_F+m_{F'})$.
For example, when using $^{87}$Rb bosons to realize the Hamiltonian, one can select $|\!\uparrow\rangle=|F=2,m_F=-2\rangle$ and $|\!\downarrow\rangle=|1,-1\rangle$, which gives $\lambda=\lambda_{\rm min}=1/3$.

{\em Discussion and conclusion.---}We have predicted for the first time a new quantum phase with coexisting localized, extended, and critical zones in a quasiperiodic optical Raman lattice model which is of high experimental feasibility, and uncovered the underlying mechanism to induce this phase.
We introduce the extended-state fraction $N_e/N$, critical-state fraction $N_c/N$, and localized-state fraction $N_l/N$ to distinguish the different characteristic zones. The fractal dimensions of extended, critical, and localized states 
exhibit sharp contrast in size-dependence, from which 
the energy windows and number of states of the different zones can be determined. Accordingly, this new quantum phase can be detected in experiment by measuring the momentum distribution of eigenstates and transport behaviors. Finally, we proposed and studied in detail the experimental realization of the current prediction with ultracold bosons and fermions.

This work adds a new family member to the fundamental phases in disordered systems, enriches the localization phenomena and extends the concept of mobility edges, with many interesting issues deserving further in-depth study. In particular, whether the three different zones are stable when adding finite interactions, namely, is there a correlated quantum phase with coexisting many-body localized, critical and ergodic zones? Further, can the similar quantum phase exist in the higher dimensions and in real materials? These issues are fundamentally important and should be explored in next studies.

\begin{acknowledgments}
This work is supported by National Key Research and Development Program of China (2021YFA1400900), the National Natural Science Foundation of China (Grants No.11825401, No.12104205), and the Open Project of Shenzhen Institute of Quantum Science and Engineering (Grant No. SIQSE202003). 
L. Z. acknowledges support from the startup grant of Huazhong University of Science and Technology (Grant No. 3004012191). W. S. is supported by the Key-Area Research and Development Program of Guangdong Province (Grant No.2020B0303010001, Grant No.2019ZT08X324).
\end{acknowledgments}



\global\long\def\id{\mathbbm{1}}
\global\long\def\ui{\mathbbm{i}}
\global\long\def\ud{\mathrm{d}}

\setcounter{equation}{0} \setcounter{figure}{0}
\setcounter{table}{0} 
\renewcommand{\theparagraph}{\bf}
\renewcommand{\thefigure}{S\arabic{figure}}
\renewcommand{\theequation}{S\arabic{equation}}

\onecolumngrid
\flushbottom
\newpage
\section*{\large Supplementary Material:\\Quantum phase with coexisting localized, extended, and critical zones}
In the Supplementary Materials, we first give a brief introduction to non-ergodic critical states. Then, we consider a few values of $\lambda$ in the Hamiltonian, study the underlying mechanism for the occurrence of the coexisting phase, and perform the finite size analysis for $N_e/N, N_c/N, N_l/N$ and the scaling index. Finally, we provide details of calculating the expression of $\lambda$ and contrasting between the lattice Hamiltonian and tight-binding model.

For convenience, we here rewrite the Hamiltonian:
\begin{eqnarray}\label{hamS}
H_0 &=& -t_0\sum_{\langle i,j\rangle}(c^{\dagger}_{i,\uparrow}c_{j,\uparrow}-c^{\dagger}_{i,\downarrow}c_{j,\downarrow})
+t_{so} [\sum_i(c^{\dagger}_{i,\uparrow}c_{i+1,\downarrow}-c^{\dagger}_{i,\uparrow}c_{i-1,\downarrow})+H.c.]\nonumber\\
 &+& \sum_{i}\delta_i(n_{i,\uparrow}-n_{i,\downarrow})+ \lambda\sum_{i}\delta_i(n_{i,\uparrow}+n_{i,\downarrow}),
\end{eqnarray}
where $\delta_j=M_Z\cos(2\pi\alpha j)$. We also set $t_0=1$ and $\alpha=(\sqrt{5}-1)/2$ under open boundary conditions (OPC), which is approximated by $\alpha=\frac{F_{m-1}}{F_m}$ with $F_m$ being the Fibonacci numbers under periodic boundary conditions (PBC).

\section{I. A brief introduction to non-ergodic critical states}
In the main text, we have mentioned that critical phases are fundamentally different from the extended and localized phases in the spectral statistics, multifractal properties of wave functions, dynamical evolutions, and so on, as summarized in Table Table~\ref{wavecri}.

\begin{table}[h]\renewcommand{\arraystretch}{1.5}
  \centering
\begin{tabular}{|c|c|c|c|ll}
\cline{1-4}
   \ phases \ &  \ energy spectrum \ & \ eigenstate's distribution \ & \ dynamical evolution \ & \\ \cline{1-4}
   \ extended \ & \ absolutely continuous\ &\ ergodic & ballistic & \\ \cline{1-4}
   \ localized \ & point spectrum\ & \ exponentially localized \ &\ localized\ &\\ \cline{1-4}
   \ critical \ & singular continuous & multifractal & diffusive & \\ \cline{1-4}
\end{tabular}
\caption{The significant differences between the critical phase and extended and localized phases in one dimensional systems.}\label{wavecri}
\end{table}

According to a large amount of papers, the critical states are extended but non-ergodic~\cite{Kravtsov2015S,XiaopengLi2016S,Kravtsov2018S,Deng2019S,Whitlock2019S,Das2022S}. The extended property can be easily seen from a number of perspectives, such as the fractal dimension and scaling index. To obtain an intuitive understanding, we consider the scaling index,
\begin{equation}\label{eta}
\eta_j=-\frac{\ln n_j}{\ln L}
\end{equation}
for every lattice site, where $n_j=n_{j,\uparrow}+n_{j,\downarrow}$ is the probability measure at the site $j$. We just focus on the minimal index $\eta_{min}$, which corresponds to the distribution peak $n_{max}$. For localized, extended and critical states, $\eta_{min}$ tends to $0$, $1$ and $0<\eta_{min}<1$ when $L\rightarrow\infty$, respectively.  As Anderson's definition of localization, a particle is initially localized at a site, the probability of the particle being found at this site is not zero after a long enough time. Thus, $n_{max}$ should be finite and size-independent. For the critical state, $n_{max}$ tends to zero when $L\rightarrow\infty$ and thus, critical states are delocalized. Besides exponentially localized states, there also exist the algebraically localized states,
which have a pronounced peak with the tails decaying algebraically, instead of exponentially~\cite{Saha1907S}. The localization length of algebraically localized states may be infinite, which is kind of similar with the extended or critical states. However, the algebraically localized states have a peak $n_{max}$, which is size-independent, and thus, as Anderson's definition, they are localized. Thus, we note that the critical states are different from the algebraically localized states.

\begin{figure}[h]
\includegraphics[width=0.4\textwidth]{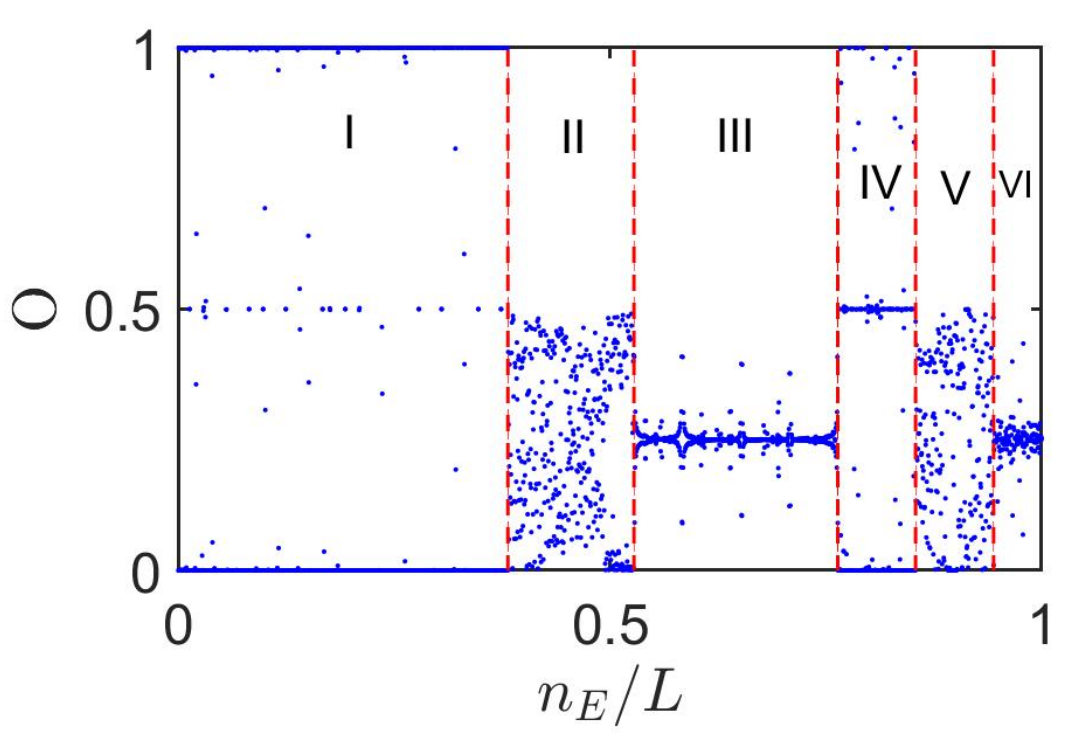}
\caption{\label{S0}
  The observable $O(E)$ as a function of $n_E/L$, where $n_E$ is the index of eigenvalues in ascending order. Here $t=1$, $t_{so}=0.5$ and $\lambda=1/3$ and $L=F_{17}=2584$ under PBC.}
\end{figure}

On the other hand, the critical state is non-ergodic~\cite{Kravtsov2015S,XiaopengLi2016S,Kravtsov2018S,Deng2019S,Whitlock2019S,Das2022S}. The non-ergodic property can also be understood from the above scaling index. We see that the distributions in all sites tend to $0$ as $L^{-1}$ for extended states, while for critical states, the distributions tends to $0$ as $L^{-\eta_j}$ with $\eta_j$ being site-dependent. To clearly describe the non-ergodic properties, we further consider the observable $O(E)=\sum_{j=1}^{L/4}\langle\psi_{E}|n_{j,\uparrow}+n_{j,\downarrow}|\psi_{E}\rangle$ quantifying the distribution in a quarter of the lattice for the eigenstate $|\psi_{E}\rangle$ with eigenvalue $E$. The fluctuations of $O(E)$ among nearby eigenstates are large (small) for the non-ergodic (ergodic) states. Fig.~\ref{S0} shows the $O(E)$ as a function of $n_E/L$.
We can see that in the ergodic zones (zones III and VI), the fluctuations of $O$ are small and near $0.25$, which is due to that we define $O(E)$ by using a quarter of the system. In
the localized zones (zones I and IV) and critical zones (zones II and V), the fluctuations are larger, meaning that they are non-ergodic. These descriptions that based on the scaling index and the observable $O(E)$ are consistent with the statement about the non-ergodic property of the critical state in previous articles~\cite{Kravtsov2015S,XiaopengLi2016S,Kravtsov2018S,Deng2019S,Whitlock2019S,Das2022S}.

\section{II. Other values of $\lambda$}
In the main text, we have discussed the case of $\lambda=1/3$. In this section, we discuss other values of $\lambda$.

\begin{figure}[b]
\includegraphics[width=1\textwidth]{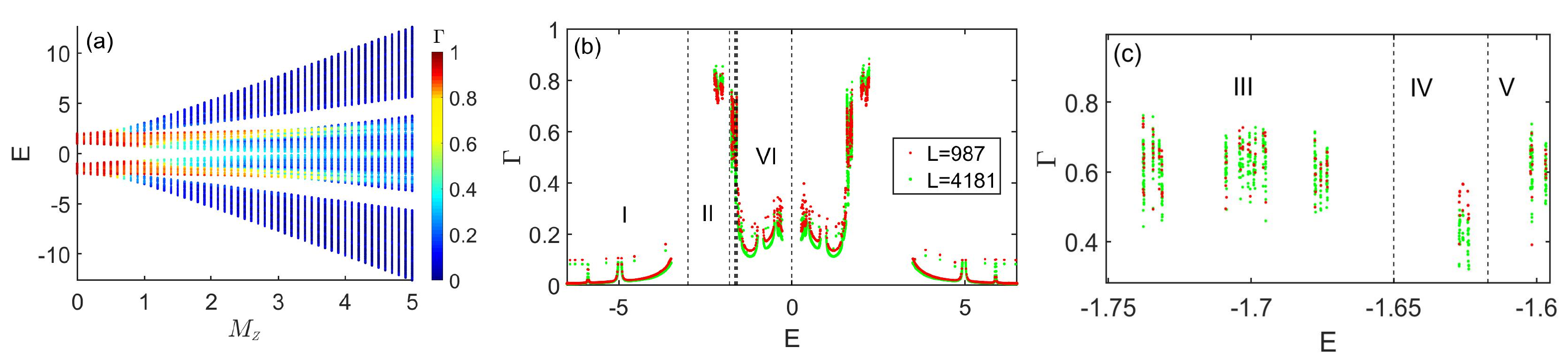}
\caption{\label{S1}
  (a) Fractal dimension $\Gamma$ of different eigenstates as the function of the corresponding eigenvalues $E$ and $M_Z$ with size $L=F_{14}=610$. (b) $\Gamma$ of different eigenstates versus the corresponding eigenvalues $E$ for different sizes $L=F_{15}=987$ and $L=F_{18}=4181$ with fixed $M_Z=2.5$. (c) The enlarged part of the zones $III, IV, V$ in Fig.(b). Other parameters are $t_{so}=0.5$ and $\lambda=1.5$ under PBC.}
\end{figure}

\subsection{A. $\lambda=1.5$ and $\lambda=0.05$}

\begin{figure}
\includegraphics[width=0.8\textwidth]{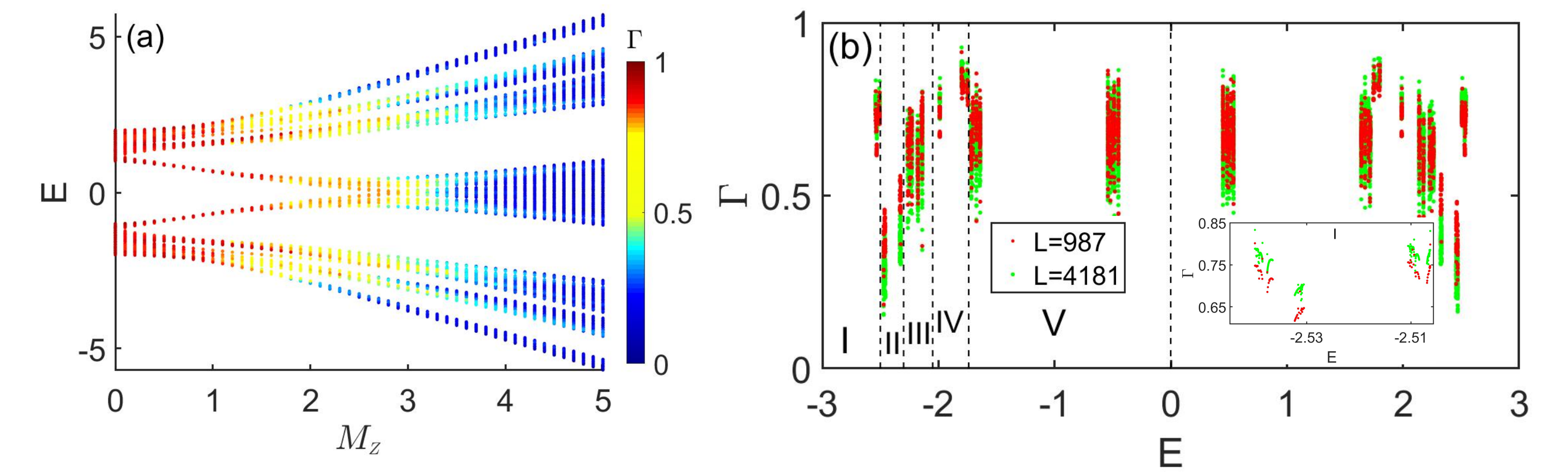}
\caption{\label{S2}
  (a) Fractal dimension $\Gamma$ of eigenstates as the function of the corresponding eigenvalues $E$ and $M_Z$ with size $L=610$. (b) Fractal dimension $\Gamma$ versus eigenvalues $E$ for different sizes $L=987$ and $L=4181$ with fixed $M_Z=1.5$. Inset: the enlarged part of the zone $I$. Other parameters are $t_{so}=0.5$ and $\lambda=0.05$ under PBC.}
\end{figure}

In the main text, we have mentioned that the quantum phase with coexisting localized, extended, and critical zones can occur in a wide range of $\lambda$. To further illustrate this, we firstly consider two cases of $\lambda=1.5$ and $\lambda=0.05$, as shown in Fig.~\ref{S1} and Fig.~\ref{S2} respectively. From Fig.~\ref{S1}(b) and (c), one can observe that the fractal dimension $\Gamma$ of all states in zones I, IV and VI become small and tend to zero with increasing the system sizes, suggesting that all states in these zones are localized; all states in zone II become large and tend to one with increasing the system sizes, meaning that these states are extended; for all states in zones III and V, $\Gamma$ are size-independent, which suggests that these states are critical. By calculating the scaling index $\eta$ defined in the main text and performing the scaling analysis, we can further determine the localized properties of states in different zones and give the same conclusions.
The same analyses are applied to Fig.~\ref{S2}(b), and we obtain that all states in zones I and IV are extended, in zones III and V are critical, in zone II are localized. Between $\lambda=1.5$ and $\lambda=0.05$, numerical results clearly show that the coexisting phase occurs. When $\lambda>1.6$, the critical states account for such a low proportion of all states that the critical region becomes approximately the critical point, i.e., $N_c/N\rightarrow 0$ and then the coexisting phase disappear.
When $\lambda<0.02$, the critical states account for a very high proportion and the critical zone becomes approximately the critical phase with all states being critical. Although $\kappa=\frac{N_e}{N}\times\frac{N_c}{N}\times\frac{N_l}{N}>0$, i.e., the coexisting phase of three different zones occur when $0.02<\lambda<0.05$ and $1.5<\lambda<1.6$, one of $N_e/N$, $N_c/N$ and $N_l/N$ is small, i.e., $\kappa$ is small. Thus, the coexisting phase exists but it is not very evident. In general, one can clearly measure the three different zones between $\lambda=0.05$ and $\lambda=1.5$.

\subsection{B. $\lambda=1$}
\begin{figure}
\includegraphics[width=0.55\textwidth]{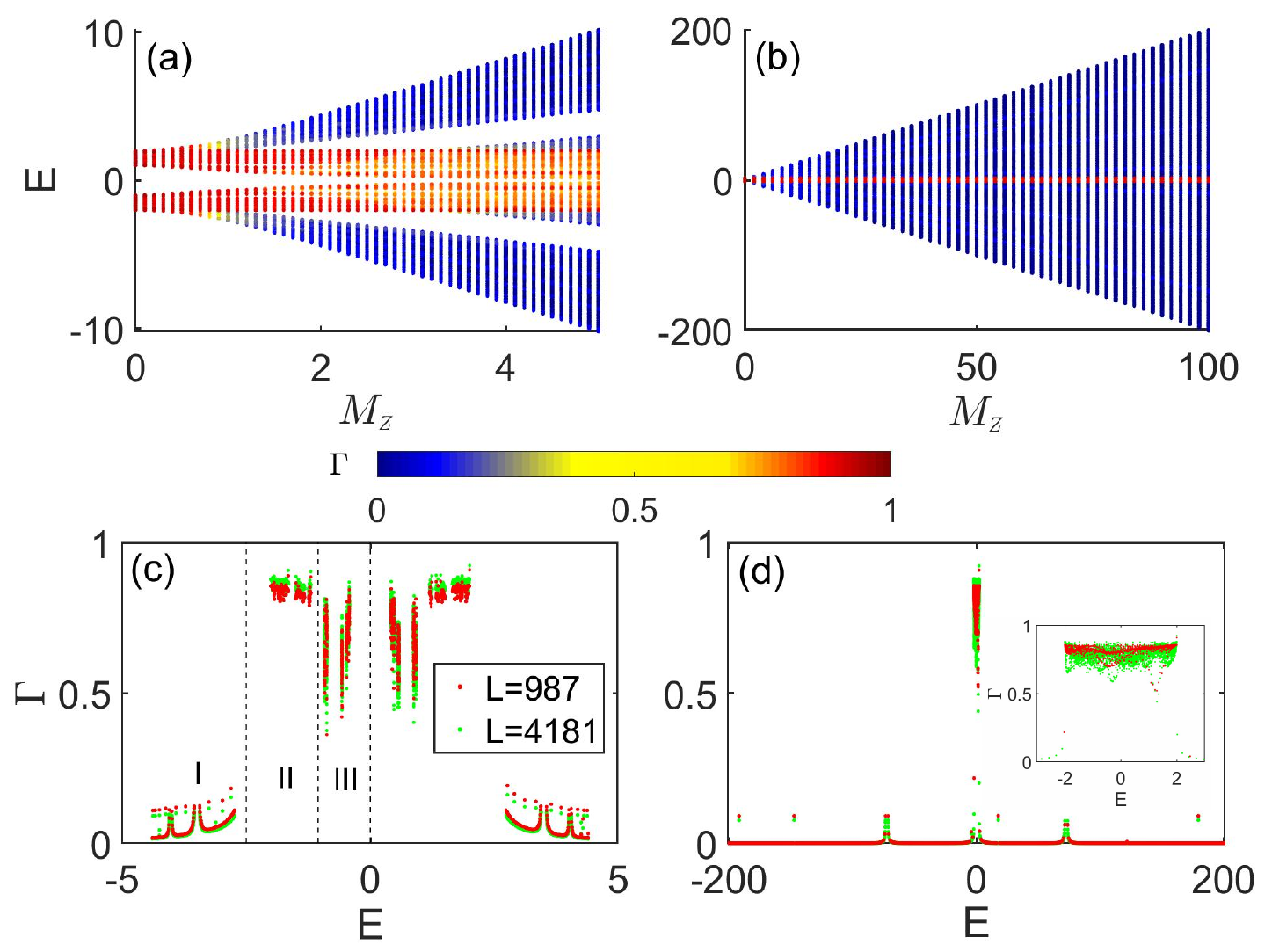}
\caption{\label{S3}
  Fractal dimension $\Gamma$ of different eigenstates as the function of the corresponding eigenvalues $E$ and $M_Z$ with $L=610$. The range of $M_Z$ is (a) from $0$ to $5$ and (b) from $0$ to $100$. $\Gamma$ versus eigenvalues $E$ for different sizes with fixed (c) $M_Z=2$ and (d) $M_Z=100$. Inset of (d): the enlarged part of the zone with energy from $-3$ to $3$. Other parameters are $t_{so}=0.5$ and $\lambda=1$ under PBC.}
\end{figure}

Next we discuss the case of $\lambda=1$, which will give an interesting physical phenomenon. From Eq.~(\ref{hamS}), when $\lambda=1$, the quasiperiodic potential applies only to the spin-up particles. Unlike other cases with $\lambda\neq 1$, there are the ever-present mobility edges at $E_c=\pm 2$ even when the quasiperiodic potential strength $M_Z$ tends to infinity, as shown in Fig.~\ref{S3}(a) and (b). Within the tailored parameter regime, one can find the coexisting phase, as shown in Fig.~\ref{S3}(c), where all states in zone $I$ are localized, all states in zone $II$ are extended and all states in zone $III$ are critical. The eigen-energies of eigenstates in zones $II$ and $III$ lie between $-2$ and $2$ and the localized states in zone $I$ corresponding energy $|E|>2$. When $M_Z$ is large, as shown in Fig.~\ref{S3}(d), all states in the energy window from $-2$ to $2$ are delocalized and all states with the energy $|E|>2$ are still localized.

\section{III. The underlying mechanism of the coexisting phase}
In this section, we discuss the underlying mechanism for the occurrence of the coexisting quantum phase. As we have known that when $\lambda=0$, the system with only quasiperiodic Zeeman potential hosts three distinct phases with solely extended, critical, or solely localized eigenstates, and the phase boundaries are $M_Z=2|t_0-t_{so}|$ between the extended and critical phases, and $M_Z=2(t_0+t_{so})$ between the critical and localized phases~\cite{Wang2SS}. The further inclusion of the spin-independent quasiperiodic potential (i.e. $\lambda\neq0$) breaks the chiral symmetry~\cite{XJLiuSS} that protects the critical phase, leading to the highly nontrivial three-zone coexisting phase associated with two new types of MEs.

\begin{figure}
\includegraphics[width=0.7\textwidth]{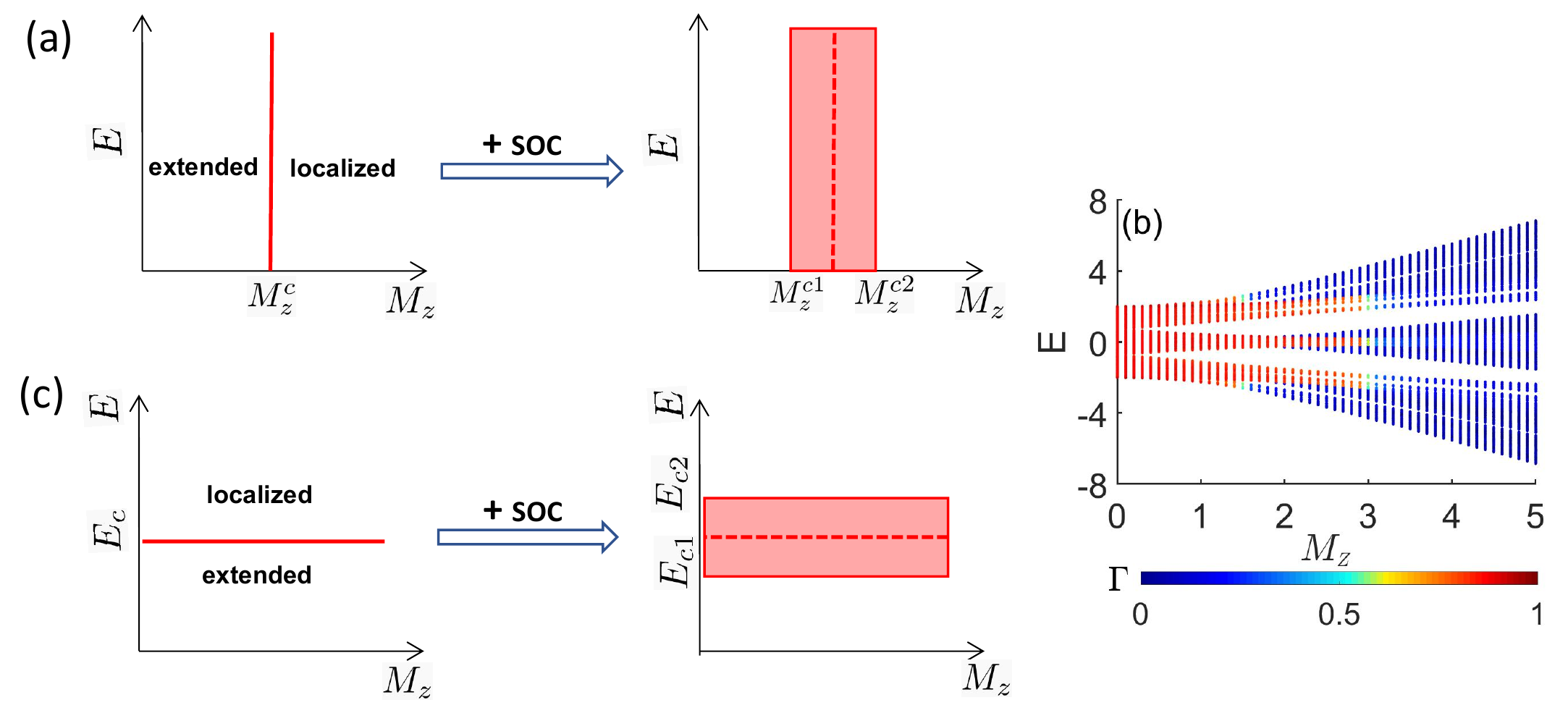}
\caption{\label{S4}
  (a) Under the action of the SOC term, the critical point $M^c_z$ become a critical region with left boundary $M^{c1}_z$ and right boundary $M^{c2}_z$. (b) Fractal dimension $\Gamma$ of different eigenstates as the function of the corresponding eigenvalues $E$ and $M_Z$ with $t_{so}=0$, $\lambda=1/3$ and $L=610$ under PBC. (c) Under the action of the SOC term, the energy-dependent critical point $E_c$ become a critical zone with lower boundary $E_{c1}$ and upper boundary $E_{c2}$.}
\end{figure}

The underlying mechanism is profound and enlightening. To see this clearly, we firstly take $t_{so}=0$, in which case the Hamiltonians for spin-up and -down components are decoupled, i.e., $H_{0}=H_{0,\uparrow}+H_{0,\downarrow}$ with
\begin{eqnarray}
 H_{0,\uparrow} = \sum_{i}[-t_0(c^{\dagger}_{i,\uparrow}c_{i+1,\uparrow}+H.c.)+ (1+\lambda)M_Z\cos(2\pi\alpha i)n_{i,\uparrow}],\label{tb3}\\
 H_{0,\downarrow} = \sum_{i}[t_0(c^{\dagger}_{i,\downarrow}c_{i+1,\downarrow}+H.c.)+ (\lambda-1)M_Z\cos(2\pi\alpha i)n_{i,\downarrow}].\label{tb4}
\end{eqnarray}
The two decoupled Hamiltonians have similar form as the AA model~\cite{AASS}, but with different quasiperiodic potential strengths.
The extended-localized phase transition point is $M^{c}_z=2|t_0/(1+\lambda)|$ for the spin-up component, and $M^{c}_z=2|t_0/(\lambda-1)|$ for the spin-down component. This leads to mobility edges. On the other hand, when $\lambda=0$ but with finite spin-orbit coupling term ($t_{\rm so}\neq 0$), the present system has a critical phase, i.e., adding the spin-orbit coupling (SOC) term ($t_{so}\neq 0$) makes the critical transition point become a critical region~\cite{Wang2SS}, as sketched in Fig.~\ref{S4}(a). When $\lambda\neq 0$, the extended-localized transition points for the spin-up and spin-down components do not coincide, for which the chiral symmetry breaking results in the appearance of mobility edges for certain range of $\lambda$, as shown in Fig.~\ref{S4}(b). Similar to the $\lambda=0$ case, the energy-dependent critical point, i.e., mobility edge $E_c$, is also extended to a critical zone if introducing the SOC term, as sketched in Fig.~\ref{S4}(c). Thus with both nonzero $\lambda$ and nonzero $t_{\rm so}$, the coexisting phase is obtained. Therefore, we conclude that the quasiperiodic Zeeman term and the spin-independent incommensurate potential, which breaks the chiral symmetry, results in the three-zone coexisting quantum phase in the presence of the SOC term.
The physical intuition is that the element inducing critical phase can drive the extended and localized states near a mobility edge into critical states, leading to a finite critical zone (while not a full critical phase) under appropriate condition.

\begin{figure}
\includegraphics[width=0.7\textwidth]{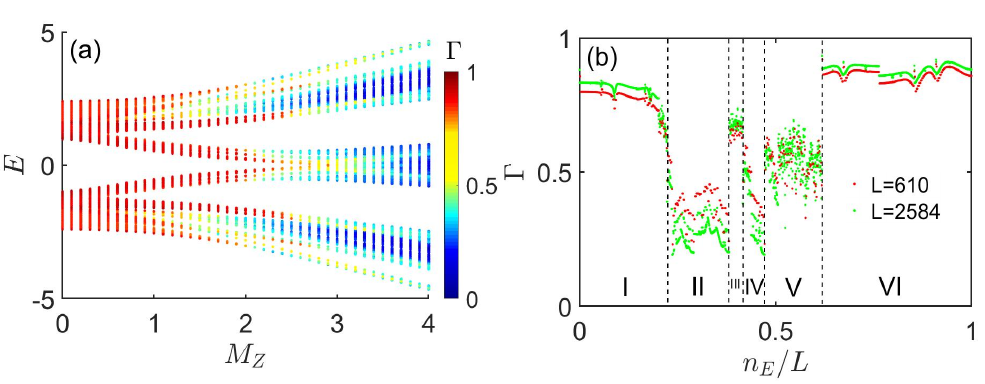}
\caption{\label{Sadd1}
  (a) Fractal dimension $\Gamma$ of different eigenstates as the function of the corresponding eigenvalues $E$ and $M_Z$ with $t_0=1$, $t_{so}=0.5$, $t_1=0.2$, and $L=610$. (b) $\Gamma$ of different eigenstates versus $n_E/L$ with $n_E$ being the corresponding eigenvalue index for different sizes $L=610$ and $L=2584$ with $M_z=1.4$.  }
\end{figure}

This mechanism is widely applicable and can give rise to broad range of systems for the realization of coexisting phase. For example, there are many schemes to obtain mobility edges, such as, the quasiperiodic systems with the next-nearest-neighbor hopping~\cite{Biddle2009S} or introducing the exponential~\cite{BiddleS} or power-law hopping~\cite{Deng2019S} term,
or adding other potentials~\cite{XLiS,HYaoS}. Combining either of them with the way to induce critical phase can yield the coexisting phase.
For the Hamiltonian~(\ref{hamS}), these is a critical phase when $\lambda=0$~\cite{Wang2SS}. Now we fix $\lambda=0$ and then add a next-nearest-neighbor hopping term $H_1=-t_1\sum_{\langle \langle i,j\rangle\rangle}(c^{\dagger}_{i,\uparrow}c_{j,\uparrow}-c^{\dagger}_{i,\downarrow}c_{j,\downarrow})$, where $\langle \langle i,j\rangle\rangle$ denotes that $i, j$ are the next-nearest-neighbor sites and $t_1$ is the strength of the next-nearest-neighbor hopping. As the discussions above, this system include the elements inducing the mobility edges and critical phase, and we can obtain the coexisting phase, as shown in Fig.~\ref{Sadd1}. From Fig.~\ref{Sadd1}(b), we see that all states in zones I, III and VI are extended, in zones II and IV are localized, and in zone V are critical. Further, we also find that if the hopping term in the Hamiltonian~(\ref{hamS}) with $\lambda=0$ becomes the long range hopping, the new coexisting phase can also be easily obtained.  Thus, in real condensed matter or cold atom systems, the new coexisting phase is more widespread and more easily realized than the pure critical phase.

\section{IV. finite size analysis}
\begin{figure}[h]
\includegraphics[width=0.75\textwidth]{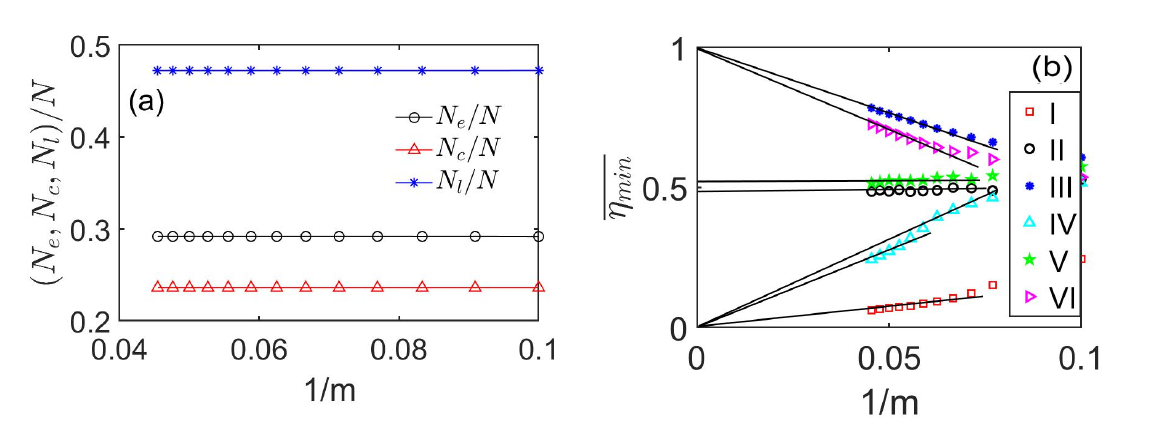}
\caption{\label{S5}
(a)$N_e/N$, $N_c/N$ and $N_l/N$ versus $1/m$, (b) $\overline{\eta_{min}}$ as a function of $1/m$ for different zones, where $m$ are the Fibonacci indices, which give the system sizes $L=F_{m}$. Other parameters are $t_{so}=0.5$, $\lambda=1/3$ and $M_Z=1.5$.}
\end{figure}

In this section, we consider the finite size effect of $N_e/N$, $N_c/N$, $N_l/N$ and $\overline{\eta_{min}}$. Fig.~\ref{S5}(a) shows $N_e/N$, $N_c/N$, $N_l/N$ versus $m$, where $m$ are the Fibonacci indices and the corresponding system sizes $L=F_{m}$, with $t_{so}=0.5$, $\lambda=1/3$ and $M_Z=1.5$. It can be seen that $N_e/N$, $N_c/N$, $N_l/N$ are invariable when the system sizes increased.

 As we have discussed in the main text,
the localized properties of all states in the same zone are same, so we can consider the average $\eta_{min}$:
\begin{equation}\label{aeta}
\overline{\eta_{min}}=\frac{1}{N_s}\sum_{same\: zone} \eta_{min},
\end{equation}
where $N_s$ is the total number of eigenstates in the same zone. Fig.~\ref{S5}(b) shows the $\overline{\eta_{min}}$ of different zone and we see that it extrapolates to $0$ in zones I and IV, to $1$ in zones III and VI, to the values far from $0$ and $1$ in zones II and V, which confirm that the corresponding eigenstates in these zones are localized, extended and critical respectively. Thus, we confirm that there can exist extended, critical and localized zones when varying the eigenvalues.

\begin{figure}[h]
\includegraphics[width=0.6\textwidth]{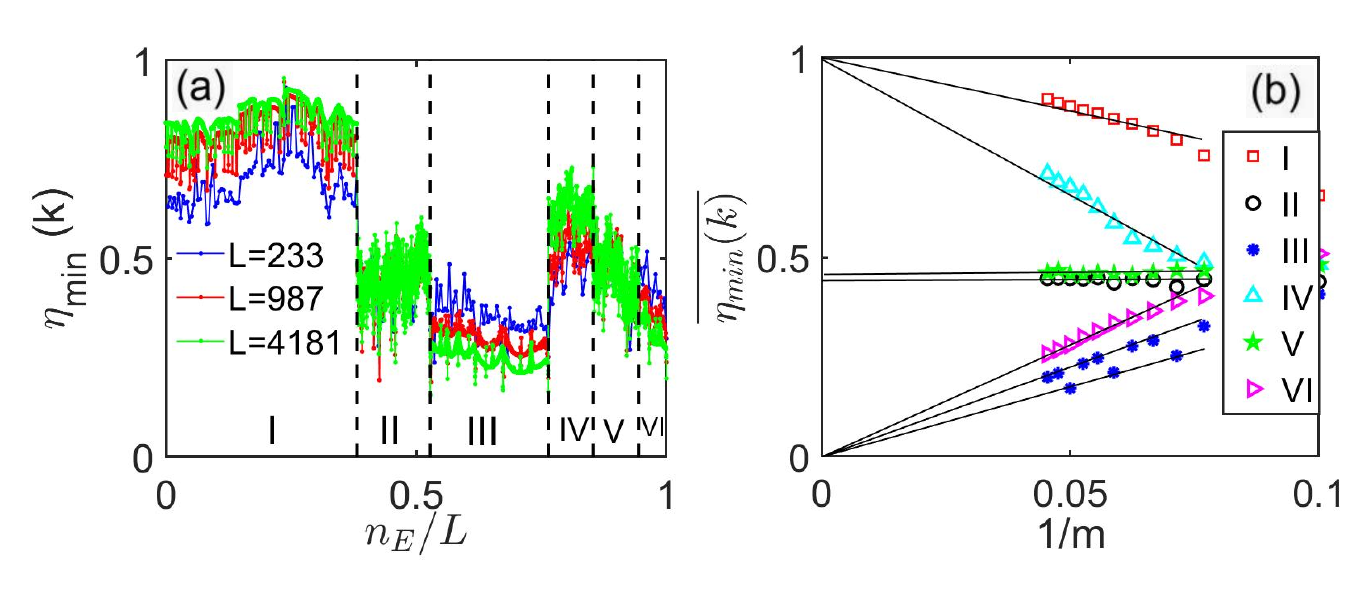}
\caption{\label{S7}
  (a) $\eta_{min}$ in momentum space for each eigenstate at three different sizes. (b) $\overline{\eta_{min}(k)}$ as a function of $1/m$ for different zones, where $m$ are the Fibonacci indices. Other parameters are $t_{so}=0.5$, $\lambda=1/3$ and $M_Z=1.5$.}
\end{figure}

We can also define a scaling index in momentum space,
\begin{equation}\label{eta}
\eta(k)=-\frac{\ln n(k)}{\ln L}
\end{equation}
for every momentum $k$. We focus on the minimal index $\eta_{min}(k)$, which corresponds to the distribution peak $n_{max}(k)$. For localized distributions in the momentum space, $n_{max}(k)$ is 
a non-zero finite value when $L\rightarrow\infty$, so $\eta_{min}(k)$ tends to zero. For extended and critical distributions, $n_{max}(k)$ tends to zero in the different ways due to their ergodic and non-ergodic, respectively. This leads to $\eta_{min}(k)=1$ and $0<\eta_{min}(k)<1$ for extended and critical states when $L\rightarrow\infty$, respectively. From the results of $\eta_{min}(k)$ [Fig.~\ref{S7}(a)], one observes a monotonic decrease (increase) with increasing the system sizes in zones III and VI (I and IV), implying that all states are localized (extended) in momentum space and the corresponding states are extended (localized) in real space. In zones II and V, $\eta_{min}(k)$ is not sensitive to the lattice sizes, meaning that the corresponding states are critical.

For the sake of experimental detection, we can also define the averaged $\eta(k)_{min}$ over the eigenstates in a single zone,
\begin{equation}\label{aeta}
\overline{\eta_{min}(k)}=\frac{1}{N_s}\sum_{same\: zone} \eta_{min}(k),
\end{equation}
where $N_s$ is the number of states in the zone. Fig.~\ref{S7} (b) shows that $\overline{\eta_{min}(k)}$ extrapolates to $1$ in zones I and IV, to $0$ in zones III and VI, while to the values far from $0$ and $1$ in zones II and V, meaning that the corresponding states in these zones are localized, extended and critical, respectively. Therefore, the corresponding eigenstates are extended, localized, and critical in real space, respectively.

\section{V. Experimental realization}

As the discussions in the main text and Refs.~\cite{Wang2SS,XJLiuSS,BSongSS,BZWangSS}, we realize the model in a 1D optical Raman lattice with incommensurate weak lattices.
The total Hamiltonian reads
\begin{equation}\label{Htotal}
H=\left[\frac{k_z^2}{2m}+V_1(z)+\lambda V_2\right]\otimes\id+{\cal M}(z)\sigma_x+\left[V_2(z)+\frac{\delta}{2}\right]\sigma_z,
\end{equation}
where $V_1(z)=V_p\cos^2(k_1z)$ is the primary lattice, $V_2(z)=\frac{V_s}{2}\cos(2{k}_2z+\phi)$ denotes the secondary spin-dependent weak lattice, giving an incommensurate Zeeman potential with the irrational number $\alpha=k_2/k_1$, ${\cal M}(z)=M_0\cos(k_1z)$ is the Raman coupling potential, and $\delta$ is the two-photon detuning.

This setup makes use of two standing waves and one plane-wave beam. With the bias field ${\bf B}$ applied along the $z$ direction, a standing-wave beam  ${\bf E}_{1}=\hat{e}_+E_{1+}\cos(k_1z)+\hat{e}_-E_{1-}\cos(k_1z)$ with $x$ polarization
and a plane wave ${\bf E}_3=\hat{e}_zE_3 e^{\ui k_3x}$ with $z$ polarization can be applied to generate the primeary lattice $V_1(z)=V_{\rm p}\cos^2({k}_1z)$ with the depth $V_{\rm p}\propto (E^2_{1+}+E^2_{1-})/\Delta$,
and a Raman coupling potential  ${\cal M}(z)=M_{0}\cos({k}_1z)$, with $M_{0}\propto E_{1-}E_3/\Delta$, simultaneously.
Here $\Delta$ denotes the coupling detuning.
The former induces the spin-conserved hopping and the latter produces the spin-flip hopping.
In addition, another standing wave ${\bf E}_2=\hat{e}_+E_{2+}\cos(k_2z+\phi/2)+\hat{e}_-E_{2-}\sin(k_2z+\phi/2)$,
formed by two counter-propagating lights with proper polarization, can be used to produce
a spin-dependent lattice $V_2(z)=\frac{V_{\rm s}}{2}\cos(2{k}_2z+\phi)$, with $V_{\rm s}\propto (E^2_{2+}-E^2_{2-})/\Delta$.
Note that the parameter $\lambda$ can vary within the range $\lambda_{\rm min}\leq \lambda<\infty$ by tuning the light polarization of the beam ${\bf E}_2$.


In the following, we consider the mutually perpendicular configuration of the beam ${\bf E}_2$ which corresponds to that the parameter $\lambda$ takes a minimal value $\lambda_{\rm min}$.
For alkali-mental atoms, the lasers configuration constructs AC Stark shift~\cite{zhang2021latticeS,2010NJPhS}
\begin{equation}
V\left(F, m_{F}\right)=-2 A\left[\alpha^{(0)}+\alpha^{(1)} m_{F}g_{F} \cos (2 k z-\phi+\pi/2)\right],
\label{acstark}
\end{equation}
with
\begin{equation}
\begin{aligned}
&\alpha^{(0)}(\omega) \approx-\frac{\pi c^{2} \Gamma_{D_2}}{2 \omega_{0}^{3}}\left(\frac{2}{\Delta_{D_2}}+\frac{1}{\Delta_{D_1}}\right), \\
&\alpha^{(1)}(\omega) \approx-\frac{\pi c^{2} \Gamma_{D_2}}{2 \omega_{0}^{3}}\left(\frac{1}{\Delta_{D_2}}-\frac{1}{\Delta_{D_1}}\right),
\end{aligned}
\end{equation}
where $A=2 \epsilon_{0} c|E|^{2}$ is laser field intensity, $\Gamma_{D_2}$ is the decay rate of the $D_2$-line exited state, $\omega_{0}$ is the central atomic transition frequency, $\Delta_{D_2} (\Delta_{D_1})$ is the laser detuning to $D_2 (D_1)$ line~\cite{Wang2SS}, $m_{F}$ is the magnetic quantum number, $F$ is the total angular momentum, $g_{F}$ is the hyper-fine structure Land\'e factor, and $k$ is the wave vector. The first term in Eq.~\eqref{acstark} is a global energy shift, while the second term is a periodic lattice potential dependent on $(F,m_{F})$. By choosing two hyperfine states $\left|F, m_{F}\right\rangle$ and $\left|F^{\prime}, m_{F}^{\prime}\right\rangle$ to construct the spin-1/2 system, the lattice potential can be written as (the first term is neglected)
\begin{equation}
\left[\begin{array}{cc}
-2A\alpha^{(1)} m_{F}g_{F} \cos (2 k z-\phi+\pi/2) & 0 \\
0 & -2A\alpha^{(1)} m_{F}^{\prime}g_{F^{\prime}} \cos (2 k z-\phi+\pi/2)
\end{array}\right]=V_{-}(z) \sigma_{z}+V_{+}(z) \sigma_{0}
\end{equation}
with $V_{\pm}(z)=-2A\left(m_{F}g_{F} \pm m_{F}^{\prime}g_{F^{\prime}} \right) \alpha^{(1)} \cos (2 k z-\phi+\pi/2)$.
Here $V_{+}(z)$ and $V_{-}(z)$ represent the spin-independent and the spin-dependent lattices, respectively, with relative strength
\begin{equation}
\lambda_{\rm min}=\frac{g_{F} m_{F}+g_{F^{\prime}} m_{F}^{\prime}}{g_{F} m_{F}-g_{F^{\prime}} m_{F}^{\prime}}.
\end{equation}

\begin{figure}[h]
\includegraphics[width=0.95\textwidth]{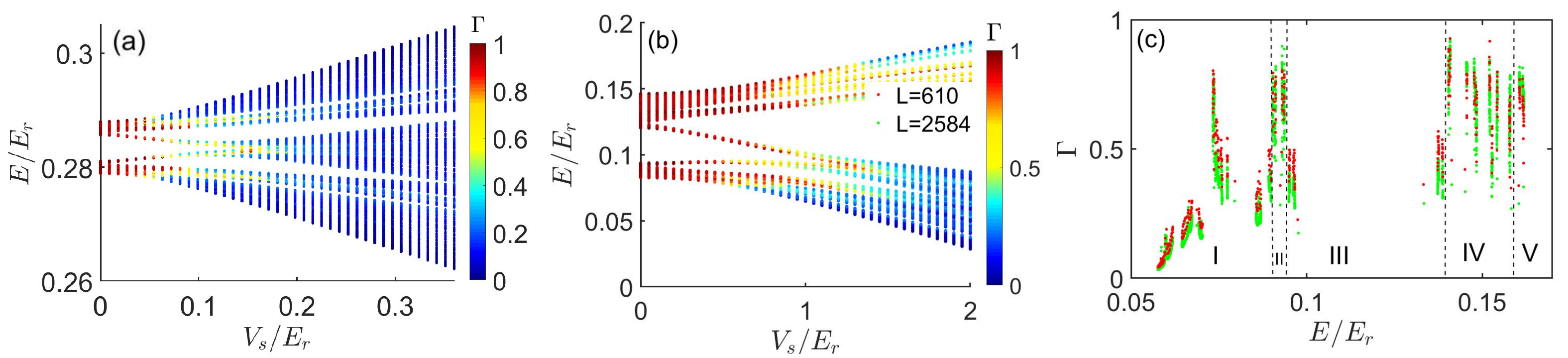}
\caption{\label{S6}
  (a) and (b) Fractal dimension $\Gamma$ of different eigenstates in the lowest $s$-bands of the lattice model as the function of the corresponding eigenvalues $E/E_r$ and the lattice depth $V_s/V_r$. Here we set $k_2=\frac{\sqrt{5}-1}{2}k_1$, $\phi=0$ and (a) $V_p=10E_r$, $M_0=1.16E_r$ with $E_r=\frac{\hbar^2k^2}{2m}$; (b) $V_p = 2Er$, $M_0 = 0.5Er$, $\lambda = 1/8$. (c) $\Gamma$ versus eigenvalues $E/E_r$ for different sizes with fixed $V_s/E_r=1.15$ and other parameters being the same with (b), and there exist three different zones: localized zones (I and III), critical zones (II and IV) and extended zone (V).}
\end{figure}

The continuous Hamiltonian~(\ref{Htotal}) indeed leads to the model Hamiltonian~(\ref{hamS}) in the tight-binding limit. To further verify the realization scheme, we also calculate the fractal dimension, defined in the main text, of the lowest $s$-band eigenstates of the continuous  Hamiltonian~(\ref{Htotal}), as shown in Fig.~\ref{S6}(a). It can be seen that the fractal dimensions of eigenstates are very similar to the results in Fig.3(a) of the main text. However, the range of the critical zone is less than
$0.01E_r$, and thus, in experiment, one need very high energy resolution, i.e., very low temperature, to distinguish it. To obtain a bigger critical zone, we make the lattice become shallow and reduce $\lambda$,
because the critical zone becomes a critical phase when $\lambda = 0$. The lattice become shallow, which will introduce the next-nearest-neighbor hopping and even long-range hopping, and as discussed above, it does not make the coexisting phase disappear.
Fig.~\ref{S6}(b) shows the fractal dimension of different eigenstates with $V_p = 2Er$, $M_0 = 0.5Er$ and $\lambda = 1/8$. One can find that the energy ranges of different zones become bigger and can reach $0.03E_r$ [see Fig.~\ref{S6}(c)]. Thus, by making the lattice become shallow and reducing $\lambda$, one can obtain bigger distinguishable zones, which make it easier to detect different zones, because the required energy resolution is less.

\begin{thebibliography}{70}
\bibitem{explain0} In mathematical terminology, the energy spectrum is absolutely continuous and singular continuous in extended and critical phases, respectively. The localized system possesses a point spectrum~\cite{Jitomirskaya}.
\bibitem{Fibonacci} Fibonacci numbers are defined recursively by $F_{m+1}=F_{m}+F_{m-1}$ with $F_0=F_1=1$.
\bibitem{explain} In our work, we provide two different approaches to determine the number of eigenstates in different zones. The first is obtained in three steps: (i) To calculate the fractal dimensions of all eigenstates for different sizes; (ii) To determine the energy windows of different zones; (iii) To numerically compute the number of eigenstates in these different energy windows, giving $N_e/N, N_c/N, N_l/N$. Another approach is to use the ratio of the index of energy mode $n_E$ to $L$ as the the horizontal axis, instead of the eigenvalue $E$, which can also directly determine $N_e/N, N_c/N, N_l/N$. From the first approach, one can obtain the eigen-energy distribution and energy range of different zones, but sometimes one need enlarge some zones of difficult judgment to make the zone division more precise. From the second approach, one can directly obtain $N_e/N, N_c/N, N_l/N$, because there exist the sudden change between different zones (see Fig.3(c)), then one can easily obtain the energy gap of different zones numerically. Combining the two approaches, one can obtain the complete information of energy spectrum and eigenstates. The two approaches can be applied to other generic models.
\bibitem{explain2} If a zone include the mobility edge separating extended and localized states, the average $\Gamma$ does not tend to $0$ or $1$, but they typically monotonously increase or decrease with increasing the system size, depending on the relative number of extended and localized states, which is different from the critical zone. One can detect the average $\Gamma$ in any two small sub-zone to further distinguish them. If the two small sub-zone show the same scaling behavior with the zone, this zone is critical. If at least one shows the same scaling behavior with the extended or localized zone, this zone has mobility edge.
\bibitem{gamma} The value of $\gamma$ depends on concrete systems and can be determined in experiments.
\end{thebibliography}
\end{document}